\batchmode
\documentstyle[12pt]{article}
\newcommand{\nc}{\newcommand}
\nc{\renc}{\renewcommand}

\nc{\half}{{\textstyle{1\over2}}}
\nc{\etal}{\mbox{\it et al. }}
\nc{\ie}{{\it i.e.}}
\nc{\eg}{{\it e.g.}}

\renc{\thefootnote}{\arabic{footnote}}
\nc{\capt}[1]{{\bf Figure.} {\small\sl #1}}

\nc{\eqs}[2]{\mbox{Eqs.~(\ref{#1},\,\ref{#2})}}
\nc{\eq}[1]{\mbox{Eq.~(\ref{#1})}}

\nc{\figs}[2]{\mbox{Figs.~(\ref{#1},\,\ref{#2})}}
\nc{\fig}[1]{\mbox{Fig~.(\ref{#1})}}

\nc{\tag}[1]{\label{#1} \marginpar{{\footnotesize #1}}}
\nc{\mtag}[1]{\label{#1} \mbox{\marginpar{{\footnotesize #1}}}}
\renc{\baselinestretch}{1.5}
\newlength{\overeqskip}
\newlength{\undereqskip}
\nc{\be}[1]{\begin{equation} \mbox{$\label{#1}$}}
\nc{\bea}[1]{\begin{eqnarray} \mbox{$\label{#1}$}}
\nc{\Section}[2]{\section{#2}\label{#1}}
\nc{\Bibitem}[1]{\bibitem{#1}}
\nc{\Label}[1]{\label{#1}}

\nc{\eea}{\vspace{\undereqskip}\end{eqnarray}}
\nc{\ee}{\vspace{\undereqskip}\end{equation}}
\nc{\bdm}{\begin{displaymath}}
\nc{\edm}{\end{displaymath}}
\nc{\dpsty}{\displaystyle}
\nc{\bc}{\begin{center}}
\nc{\ec}{\end{center}}
\nc{\ba}{\begin{array}}
\nc{\ea}{\end{array}}
\nc{\bab}{\begin{abstract}}
\nc{\eab}{\end{abstract}}
\nc{\btab}{\begin{tabular}}
\nc{\etab}{\end{tabular}}
\nc{\bit}{\begin{itemize}}
\nc{\eit}{\end{itemize}}
\nc{\ben}{\begin{enumerate}}
\nc{\een}{\end{enumerate}}
\nc{\bfig}{\begin{figure}}
\nc{\efig}{\end{figure}}
\nc{\arreq}{&\!=\!&}
\nc{\arrmi}{&\!-\!&}
\nc{\arrpl}{&\!+\!&}
\nc{\arrap}{&\!\!\!\approx\!\!\!&}
\nc{\non}{\nonumber\\*}
\nc{\align}{\!\!\!\!\!\!\!\!&&}

\def\lsim{\; \raise0.3ex\hbox{$<$\kern-0.75em
      \raise-1.1ex\hbox{$\sim$}}\; }
\def\gsim{\; \raise0.3ex\hbox{$>$\kern-0.75em
      \raise-1.1ex\hbox{$\sim$}}\; }
\nc{\DOT}{\hspace{-0.08in}{\bf .}\hspace{0.1in}}
\nc{\Laada}{\hbox {$\sqcap$ \kern -1em $\sqcup$}}
\nc\loota{{\scriptstyle\sqcap\kern-0.55em\hbox{$\scriptstyle\sqcup$}}}
\nc\Loota{{\sqcap\kern-0.65em\hbox{$\sqcup$}}}
\nc\laada{\Loota}
\nc{\qed}{\hskip 3em \hbox{\BOX} \vskip 2ex}

\nc{\real}{{\rm I \! R}}
\nc{\Z}{{\sf Z \!\!\! Z}}
\nc{\complex}{{\rm C\!\!\! {\sf I}\,\,}}
\def\bigid{\leavevmode\hbox{\small1\kern-3.8pt\normalsize1}}
\def\id{\leavevmode\hbox{\small1\kern-3.3pt\normalsize1}}
\nc{\slask}{\!\!\!/}
\nc{\bis}{{\prime\prime}}
\nc{\pa}{\partial}
\nc{\na}{\nabla}
\nc{\ra}{\rangle}
\nc{\la}{\langle}
\nc{\goto}{\rightarrow}
\nc{\swap}{\leftrightarrow}

\nc{\EE}[1]{ \mbox{$\cdot10^{#1}$} }
\nc{\abs}[1]{\left|#1\right|}
\nc{\at}[2]{\left.#1\right|_{#2}}
\nc{\norm}[1]{\|#1\|}
\nc{\abscut}[2]{\Abs{#1}_{\scriptscriptstyle#2}}
\nc{\vek}[1]{{\rm\bf #1}}
\nc{\integral}[2]{\int\limits_{#1}^{#2}}
\nc{\inv}[1]{\frac{1}{#1}}
\nc{\dd}[2]{{{\partial #1}\over{\partial #2}}}
\nc{\ddd}[2]{{{{\partial}^2 #1}\over{\partial {#2}^2}}}
\nc{\dddd}[3]{{{{\partial}^2 #1}\over
        {\partial #2 \partial #3}}}
\nc{\dder}[2]{{{d #1}\over{d #2}}}
\nc{\ddder}[2]{{{d^2 #1}\over{d {#2}^2}}}
\nc{\dddder}[3]{{d^2 #1}\over
        {d #2 d #3}}
\nc{\dx}[1]{d\,^{#1}x}
\nc{\dy}[1]{d\,^{#1}y}
\nc{\dz}[1]{d\,^{#1}z}
\nc{\dl}[1]{\frac{d\,^{#1}l}{(2\pi)^{#1}}}
\nc{\dk}[1]{\frac{d\,^{#1}k}{(2\pi)^{#1}}}
\nc{\dq}[1]{\frac{d\,^{#1}q}{(2\pi)^{#1}}}

\nc{\cc}{\mbox{$c.c.$ }}
\nc{\hc}{\mbox{$h.c.$ }}
\nc{\cf}{cf.\ }
\nc{\erfc}{{\rm erfc}}
\nc{\Tr}{{\rm Tr\,}}
\nc{\tr}{{\rm tr\,}}
\nc{\pol}{{\rm pol}}
\nc{\sign}{{\rm sign}}
\nc{\bfT}{{\bf T }}

\def\GeV{{\rm\ GeV}}

\def\TeV{{\rm\ TeV}}

\nc{\cA}{{\cal A}}
\nc{\cB}{{\cal B}}
\nc{\cD}{{\cal D}}
\nc{\cE}{{\cal E}}
\nc{\cG}{{\cal G}}
\nc{\cH}{{\cal H}}
\nc{\cL}{{\cal L}}
\nc{\cO}{{\cal O}}
\nc{\cT}{{\cal T}}
\nc{\cN}{{\cal N}}
\nc{\rvac}[1]{|{\cal O}#1\rangle}
\nc{\lvac}[1]{\langle{\cal O}#1|}
\nc{\rvacb}[1]{|{\cal O}_\beta #1\rangle}
\nc{\lvacb}[1]{\langle{\cal O}_\beta #1 |}
\nc{\bb}{\bar{\beta}}
\nc{\bt}{\tilde{\beta}}
\nc{\ctH}{\tilde{\cal H}}
\nc{\chH}{\hat{\cal H}}
\nc{\1}{\aa}
\nc{\2}{\"{a}}
\nc{\3}{\"{o}}
\nc{\4}{\AA}
\nc{\5}{\"{A}}
\nc{\6}{\"{O}}
\nc{\al}{\alpha}
\nc{\g}{\gamma}
\nc{\Del}{\Delta}
\nc{\e}{\epsilon}
\nc{\eps}{\epsilon}
\nc{\lam}{\lambda}
\nc{\om}{\omega}
\nc{\Om}{\Omega}
\nc{\ve}{\varepsilon}
\nc{\mn}{{\mu\nu}}
\nc{\k}{\kappa}
\nc{\vp}{\varphi}

\nc{\advp}[3]{{\it  Adv.\ in\ Phys.\ }{{\bf #1} {(#2)} {#3}}}
\nc{\annp}[3]{{\it  Ann.\ Phys.\ (N.Y.)\ }{{\bf #1} {(#2)} {#3}}}
\nc{\apl}[3]{{\it  Appl. Phys. Lett. }{{\bf #1} {(#2)} {#3}}}
\nc{\apj}[3]{{\it  Ap.\ J.\ }{{\bf #1} {(#2)} {#3}}}
\nc{\apjl}[3]{{\it  Ap.\ J.\ Lett.\ }{{\bf #1} {(#2)} {#3}}}
\nc{\app}[3]{{\it Astropart.\ Phys.\ }{{\bf #1} {(#2)} {#3}}}
\nc{\cmp}[3]{{\it  Comm.\ Math.\ Phys.\ }{{ \bf #1} {(#2)} {#3}}}
\nc{\cqg}[3]{{\it  Class.\ Quant.\ Grav.\ }{{\bf #1} {(#2)} {#3}}}
\nc{\epl}[3]{{\it  Europhys.\ Lett.\ }{{\bf #1} {(#2)} {#3}}}
\nc{\ijmp}[3]{{\it Int.\ J.\ Mod.\ Phys.\ }{{\bf #1} {(#2)} {#3}}}
\nc{\ijtp}[3]{{\it Int.\ J.\ Theor.\ Phys.\ }{{\bf #1} {(#2)} {#3}}}
\nc{\jmp}[3]{{\it  J.\ Math.\ Phys.\ }{{ \bf #1} {(#2)} {#3}}}
\nc{\jpa}[3]{{\it  J.\ Phys.\ A\ }{{\bf #1} {(#2)} {#3}}}
\nc{\jpc}[3]{{\it  J.\ Phys.\ C\ }{{\bf #1} {(#2)} {#3}}}
\nc{\jap}[3]{{\it J.\ Appl.\ Phys.\ }{{\bf #1} {(#2)} {#3}}}
\nc{\jpsj}[3]{{\it J.\ Phys.\ Soc.\ Japan\ }{{\bf #1} {(#2)} {#3}}}
\nc{\lmp}[3]{{\it Lett.\ Math.\ Phys.\ }{{\bf #1} {(#2)} {#3}}}
\nc{\mpl}[3]{{\it  Mod.\ Phys.\ Lett.\ }{{\bf #1} {(#2)} {#3}}}
\nc{\ncim}[3]{{\it  Nuov.\ Cim.\ }{{\bf #1} {(#2)} {#3}}}
\nc{\np}[3]{{\it  Nucl.\ Phys.\ }{{\bf #1} {(#2)} {#3}}}
\nc{\npps}[3]{{\it  Nucl.\ Phys.\ Proc.\ Suppl.\ }{{\bf #1} {(#2)} {#3}}}
\nc{\pr}[3]{{\it Phys.\ Rev.\ }{{\bf #1} {(#2)} {#3}}}
\nc{\pra}[3]{{\it  Phys.\ Rev.\ A\ }{{\bf #1} {(#2)} {#3}}}
\nc{\prb}[3]{{\it  Phys.\ Rev.\ B\ }{{{\bf #1} {(#2)} {#3}}}}
\nc{\prc}[3]{{\it  Phys.\ Rev.\ C\ }{{\bf #1} {(#2)} {#3}}}
\nc{\prd}[3]{{\it  Phys.\ Rev.\ D\ }{{\bf #1} {(#2)} {#3}}}
\nc{\prl}[3]{{\it Phys.\ Rev.\ Lett.\ }{{\bf #1} {(#2)} {#3}}}
\nc{\pl}[3]{{\it  Phys.\ Lett.\ }{{\bf #1} {(#2)} {#3}}}
\nc{\prep}[3]{{\it Phys.\ Rep.\ }{{\bf #1} {(#2)} {#3}}}
\nc{\prsl}[3]{{\it Proc.\ R.\ Soc.\ London\ }{{\bf #1} {(#2)} {#3}}}
\nc{\ptp}[3]{{\it  Prog.\ Theor.\ Phys.\ }{{\bf #1} {(#2)} {#3}}}
\nc{\ptps}[3]{{\it  Prog\ Theor.\ Phys.\ suppl.\ }{{\bf #1} {(#2)} {#3}}}
\nc{\physa}[3]{{\it  Physica\ A\ }{{\bf #1} {(#2)} {#3}}}
\nc{\physb}[3]{{\it  Physica\ B\ }{{\bf #1} {(#2)} {#3}}}
\nc{\phys}[3]{{\it Physica\ }{{\bf #1} {(#2)} {#3}}}
\nc{\rmp}[3]{{\it  Rev.\ Mod.\ Phys.\ }{{\bf #1} {(#2)} {#3}}}
\nc{\rpp}[3]{{\it Rep.\ Prog.\ Phys.\ }{{\bf #1} {(#2)} {#3}}}
\nc{\sjnp}[3]{{\it Sov.\ J.\ Nucl.\ Phys.\ }{{\bf #1} {(#2)} {#3}}}
\nc{\spjetp}[3]{{\it Sov.\ Phys.\ JETP\ }{{\bf #1} {(#2)} {#3}}}
\nc{\yf}[3]{{\it Yad.\ Fiz.\ }{{\bf #1} {(#2)} {#3}}}
\nc{\zetp}[3]{{\it Zh.\ Eksp.\ Teor.\ Fiz.\  }{{\bf #1}  {(#2)} {#3}}}
\nc{\zp}[3]{{\it Z.\ Phys.\ }{{\bf #1} {(#2)} {#3}}}
\nc{\ibid}[3]{{\sl ibid.\ }{{\bf #1} {#2} {#3}}}
\nc{\rf}[1]{(\ref{#1})}
\nc{\nn}{\nonumber \\*}
\nc{\bfB}{\bf{B}}
\nc{\bfv}{\bf{v}}
\nc{\bfx}{\bf{x}}
\nc{\bfy}{\bf{y}}
\nc{\vx}{\vec{x}}
\nc{\vy}{\vec{y}}
\nc{\oB}{\overline{B}}
\nc{\oI}{\overline{I}}
\nc{\oR}{\overline{R}}
\nc{\rar}{\rightarrow}
\nc{\ti}{\times}
\nc{\slsh}{\hskip-5pt/}
\nc{\sm}{Standard~Model~}
\nc{\MP}{M_{\rm Pl}}
\nc{\tp}{t_{\rm Pl}}
\nc{\ave}{\bar{E}}

\nc{\eff}{{\rm eff}}
\nc{\kk}{\vek{k}}
\nc{\pp}{{\rm p}}
\nc{\ga}{g_{a\gamma}}
\nc{\vv}{\\}
\nc{\eee}{{\bf E}}
\nc{\bbb}{{\bf B}}
\nc{\qcd}{T_{\rm QCD}}
\nc{\G}{\rm \ G}
\def\vec#1{{\bf #1}}

\def\lae{\;^{<}_{\sim} \;} \def\gae{\; ^{>}_{\sim} \;} 

\def\ell{e^{c}LL}

\begin{document}
{\title{\vskip-2truecm{\hfill {{\small \\
	\hfill \\
	}}\vskip 1truecm}
{\bf  Flat Direction Dynamics in a Non-Topological Soliton-Dominated Universe}}
{\author{
{\sc  John McDonald$^{1}$}\\
{\sl\small Dept. of Mathematical Sciences, University of Liverpool,
Liverpool L69 3BX, England}
}
\maketitle
\begin{abstract}
\noindent

                   In running mass inflation and hybrid inflation models it is possible that 
the inflaton field will fragment into non-topological solitons,
 resulting in a highly inhomogeneous post-inflation era prior to reheating. 
In supersymmetric models with a conventional homogeneous post-inflation era, the 
dynamics of flat direction scalars are determined by 
$cH^{2}$ corrections to the mass squared terms 
(where $|c| \approx 1$), coming from F-terms in the early Universe
combined with Planck-scale suppressed interactions. Here we reconsider the 
mass squared corrections for a Universe dominated by inflatonic non-topological 
solitons. We show that in this case the dynamics of a coherently oscillating 
flat direction scalar are typically the same as for the case where there is 
no significant mass squared correction, even in the vicinity
 of the non-topological solitons. 
Therefore the dynamics of flat direction scalars 
in a Universe dominated by inflatonic non-topological solitons 
are equivalent to the case $c = 0$ of a homogeneous Universe. 

\end{abstract}
\vfil
\footnoterule
{\small $^1$mcdonald@amtp.liv.ac.uk}

\thispagestyle{empty}
\newpage
\setcounter{page}{1}

\section{Introduction}

           In the Minimal Supersymmetric (SUSY) Standard Model (MSSM) \cite{mssm} 
it is known 
that there are a large number of possible flat directions formed from combinations 
of squark, slepton and Higgs fields \cite{drt}. In the early Universe it is likely 
that these flat
 directions will develop a non-zero expectation value during inflation, and will 
subsequently form Bose condensates corresponding to coherently oscillating scalar 
fields. The cosmology of the formation and decay of flat direction condensates is of
 fundamental importance to the cosmology of the MSSM and its extensions 
\cite{kmaz}. They
 may be the origin of the baryon asymmetry of the Universe via 
the Affleck-Dine (AD) mechanism \cite{ad}. Flat direction condensates 
may also account for a density of dark matter 
via the formation \cite{ks}
and late decay \cite{bbb,bbb2} of Q-balls, allowing the baryon and dark matter number 
densities to be 
related. In addition, a
 SUSY flat direction condensate could possibly account for the energy density 
perturbations responsible for structure formation, serving as a SUSY curvaton 
\cite{curv,curv2,curv3,kcurv,postma,snu,phase}. 

        The dynamics of flat direction fields in the post-inflation era are largely 
determined by corrections to the mass squared terms coming from 
SUSY breaking F-terms associated with fields which contribute
 to the energy density, $\rho$. These F-terms, 
when combined with Planck-scale suppressed interactions, result in a 
mass squared term for the flat direction scalar of the order of $|F|^{2}/M^{2}$, 
where $M = M_{Pl}/sqrt{8 \pi}$ and where $M_{Pl}$ is the Planck mass.
For a conventional post-inflation Universe with a {\it homogeneous} energy density 
due to coherent inflaton oscillations after inflation, these 
corrections are of the form $cH^{2}$ with $|c|$ of the order of 1 \cite{h2,h22}. 

              However, in important examples of SUSY inflation models it has become 
clear that the energy density in the inflaton field may not remain homogeneous after 
inflation ends. It has been shown that in SUSY running mass inflation models 
\cite{krun} it is possible that 
quantum fluctuations of the inflaton 
field will grow, resulting in formation of non-topological solitons (NTS) 
\cite{icf,cpr,iballs}. 
In addition, it has been shown that in SUSY 
hybrid inflation models \cite{tp,icf} quantum fluctuations of the inflaton 
can grow to form non-linear fragments. It possible that such fragments 
will evolve into quasi-stable inflatonic NTS (for the case of a real inflaton 
field we refer to these as 'inflaton condensate lumps' \cite{icf}).  
This has yet to be established numerically 
for fully realistic SUSY hybrid inflation models \cite{mp0}, although
 some numerical evidence for NTS formation in hybrid inflation
 models exists \cite{cpr}.  
For the case of D-term inflation models \cite{dti}, 
fragmentation is expected to occur if $\lambda \gae 0.1 g$ \cite{icf,mp1}, whilst for 
F-term inflation models \cite{fti} it is expected to always occur \cite{icf,mp1}.

               In SUSY inflation models the 
inflaton is generally a complex scalar field. As a result, it is possible for Q-balls to 
form \cite{krun}. This was demonstrated numerically for the case of a running mass
 chaotic inflation model, where it was shown that initially neutral inflaton 
condensate lumps fragment
 into Q-ball, anti-Q-ball pairs (the most stable configuration) \cite{krun}. 
SUSY hybrid inflation models also have Q-ball solutions, formed from a 
combination of a complex inflaton field and a real symmetry breaking 
scalar field \cite{mp2}. If neutral inflaton
condensate lumps form at the end of SUSY
 hybrid inflation then it is likely that these will subsequently 
fragment into inflatonic Q-ball pairs, as in the running mass case. 

     Thus it is possible that in both running mass inflation models and  
 hybrid inflation models the post-inflation 
era will be highly inhomogeneous, formed initially of inflaton condensate lumps and 
subsequently of inflatonic 
Q-balls, which eventually decay to reheat the Universe \cite{icf,kreh}. 

       In this paper we wish to discuss how SUSY flat direction
 scalar field dynamics, in particular 
the dynamics of a coherently oscillating scalar field,  
would be modified during an inflatonic NTS-dominated post-inflation era. 
We first give a qualitative discussion of how the scalar field
 dynamics might be expected to change. The energy 
density of the Universe after inflation is entirely in the form of inflaton condensate
 lumps or inflatonic 
Q-balls, with little or no energy 
outside the NTS where the inflaton field is strongly exponentially suppressed.
 Thus we might expect that 
there will effectively be no scalar mass squared corrections outside the NTS, 
since $\rho$ and $|F|^{2}$ are 
very close to zero outside the NTS. However, 
in the vicinity of the NTS 
a much larger than average inflaton energy density will occur. This suggests that the
now space-dependent inflaton energy density and
 associated F-term will induce a very large mass
 squared correction ($ \gg H^{2}$) for the flat direction field
 in the vicinity of the inflatonic
 NTS. If so, this would be expected to have a strong effect on the dynamics
 of the coherently oscillating flat direction fields. However, we will show that this 
is not the case; even at the centre of the inflatonic NTS the dynamics of a coherently
 oscillating scalar field will be shown to be essentially the same
 as in the absence of any mass squared correction. 
This is because the gradient energy term, due to the spatial distortion 
of the flat direction field in the
 vicinity of the inflatonic NTS, cancels the effect
 of the effective mass squared term 
 in the vicinity of the NTS such that the amplitude of the
 coherent oscillations is essentially unaltered. 
As a result, the dynamics of a coherently oscillating 
flat direction scalar in a NTS-dominated Universe are  
equivalent to the case of a homogeneous Universe with $c =0$.

      The paper is organised as follows. 
In Section 2 we discuss the solution of the flat direction scalar field equation  
for a coherently oscillating flat direction scalar 
in the presence of an inflatonic NTS and show that the oscillation
 amplitude is typically unaffected by the presence of the
 NTS. We also give a physical argument to support this conclusion.
 In Section 3 we consider the consequences 
for Affleck-Dine baryogenesis and SUSY curvaton dynamics. In Section 4
 we present our conclusions.

\section{Gaussian Non-Topological Solitons and Flat Direction Condensates} 

           In order to understand the effect of inflatonic NTS on
 the dynamics of a coherently oscillating 
flat direction condensate, we will consider the case of a NTS with a Gaussian profile. 
We expect that the Gaussian profile will be a reasonable approximation to 
the thick-walled NTS which are expected to form during inflaton
 condensate fragmentation. 
A particular example is the 
inflatonic Q-ball which forms at the end of running mass
 chaotic inflation \cite{krun}, which  
has the same form as the Q-balls which form 
from squark and slepton flat direction condensates during 
Affleck-Dine baryogenesis \cite{bbb2}. SUSY hybrid inflation Q-balls also have a near 
Gaussian profile for the inflaton field \cite{mp2}.

    We will consider the potential of the complex inflaton $S$ to consist in general 
of a 
mass squared term plus an attractive self-interaction term which 
allows the formation of NTS,  
\be{e1} V(S) =  m_{S}^{2} |S|^{2} + V_{int}(S)   ~.\ee
For the case of the running mass chaotic inflation model 
the interaction potential has the form \cite{krun}, 
\be{e2} V_{int}(S) = Km_{S}^{2}|S|^{2} \ln \left(\frac{|S|^{2}}{\Lambda^{2}}\right)   ~,\ee
where $K < 0$ and $\Lambda$ is a renormalization
 scale. (This has the same form as the 
potential of MSSM flat directions involving squarks \cite{bbb,bbb2}.)
$m_{S}^{2}$ is assumed to originate from soft SUSY breaking \cite{krun}. 
\eq{e2} leads to Gaussian Q-ball solutions of the
 $S$ field equation of the form \cite{bbb2} 
\be{e3} S = \frac{s(r)}{\sqrt{2}} e^{i \omega t}   ~,\ee 
where
\be{e4}   s(r) \approx s_{o}e^{-r^{2}/R^{2}}    ~,\ee
\be{e5}   R \approx \frac{\sqrt{2}}{|K|^{1/2} m_{S}}   ~,\ee
\be{e6} \omega^{2} = \omega_{o}^{2} + m_{S}^{2}(1 + K)  ~,\ee
and
\be{e7} \omega_{o}^{2} \approx 3 |K| m_{S}^{2}   ~.\ee
The values of $\omega$ and $R$ above are for the specific case of the interaction 
potential of \eq{e2}. However, the form of the Gaussian Q-ball solution 
given by \eq{e3} and \eq{e4} 
is quite general. Moreover, we expect the magnitudes of 
$\omega$ and $R$ 
to be determined in most cases 
by the same dynamical mass scale in the field equations, such that 
$\omega R$ will be typically of the order of 1. 
Thus our results should apply to thick-walled inflatonic 
Q-balls in general, with the interaction potential of 
\eq{e2} providing a specific example. 

     The initial spacing between the NTS when they first form
 will typically be of the order of the radius of the NTS \cite{icf}. 
Thus after a period of Universe expansion the spacing will be large
 compared with the NTS radius.
From the Gaussian form for the inflatonic NTS amplitude, \eq{e4}, the
 inflaton field will be exponentially 
suppressed outside an NTS, such that its value is effectively zero. 
So long as the spacing between the NTS is sufficiently 
large compared with their radius ($\delta x/R \approx 10$ will result in a suppression 
factor $e^{-100}$, where $\delta x$ is the spacing between the NTS), 
including more than one NTS will not alter the nearly zero value of the inflaton field 
and so mass squared correction outside the NTS. The 
NTS effectively do not see each other because of the exponential suppression 
of the inflaton field outside the NTS. As a result, we can study flat direction scalar
 dynamics by considering the flat direction scalar field equation in the
 background of a {\it single} NTS. 

          An important issue is the equation of motion of the flat direction scalar 
in the inhomogeneous background of a NTS-dominated
 Universe. In this case the usual 
Friedmann-Robertson-Walker metric is not strictly valid and a non-trivial
calculation of the inhomogeneous metric due to the ensemble of NTS
 is necessary to obtain the correct scalar field equation
 including the effect of the expansion of the Universe.
However, we will be concerned with the effect  
of the flat direction scalar mass term induced in the vicinity of the 
inflatonic NTS on the dynamics of a coherently oscillating flat direction field.
Since this mass term is much larger than the expansion rate, 
the effect of gravitational corrections to the scalar field equation
 of motion due to the expansion of the Universe will be negligible
(i.e. the time scale over
which the flat direction field changes due to the effect of the NTS will be small
 compared with $H^{-1}$). In addition, it has been shown that gravitational 
effects typically play no role in the NTS solution itself \cite{vilja}. 
Therefore flat space may be 
 considered when solving the field equation for the flat direction scalar.

     The potential of a SUSY flat direction scalar  
$\Phi$ is expected to purely consist of a conventional 
gravity-mediated SUSY breaking mass squared term, 
\be{e8} V(\Phi) = m_{\phi}^{2} |\Phi|^{2}    ~,\ee
where $m_{\phi} \approx 100 \GeV - 1 \TeV$.
 (We assume the coherent
 oscillations have an amplitude small enough that possible
 non-renormalizable corrections to the flat direction 
 superpotential may be neglected.)
We will consider a  
non-renormalizable Planck-scale suppressed 
interaction term between the inflaton field and flat direction field
of the form which is generally expected to arise in the low 
energy effective theory from supergravity \cite{drt,h2,h22} and 
which is not excluded by any symmetry, 
\be{e9} {\cal L}_{int} = - \frac{\lambda}{M^{2}} \int d^{4}\theta S^{\dagger}S 
\Phi^{\dagger}\Phi  \equiv - \frac{\lambda}{M^{2}} |F_{S}|^{2} \Phi^{\dagger}\Phi
~,\ee
 where $M = M_{Pl}/\sqrt{8 \pi}$ and $\lambda \approx 1$.
 (This is Equation 8 of \cite{drt}) 
$F_{S}$ is the F-term of the inflaton scalar, 
\be{e9a} |F_{S}|^{2} =    
\left(\partial_{\mu}S^{\dagger}\partial^{\mu}S +
\left|\frac{\partial W}{\partial S}\right|^{2} \right) |\Phi|^{2}  ~,\ee
where $W$ is the superpotential. 
This term is of the generic form which leads to the $cH^{2}$ 
correction to scalar mass squared terms in the case of 
a conventional homogeneous post-inflation Universe.
 However,  its effect must be reconsidered 
in the case of an inhomogeneous NTS-dominated Universe.

              In the case of a conventional homogeneous post-inflation 
Universe dominated by a coherently oscillating inflaton 
condensate we have $|F_{S}|^{2} = \dot{S}^{2}/2 + V(S) \equiv 
\rho_{S}$, where $V(S) = |\partial W/\partial S|^{2}$ is the SUSY
 inflaton potential. Therefore 
$|F_{S}|^{2}/M^{2} = \rho_{S}/M^{2} \equiv 3 H^{2}/M^{2}$, 
resulting in a mass-squared correction of the order of $H^{2}$.
$\lambda > 0$ ($< 0$) would then correspond to 
having $c < 0$ ($c > 0$) in the $cH^{2}$ term. Other possible 
Planck-scale suppressed interactions with the SUSY breaking inflaton F-term 
all lead to similar corrections to the mass-squared terms \cite{drt}. 
In the following we will consider \eq{e9} as a typical example.

      In order to study the effect of the inflatonic Q-ball (which we  
refer to as the S-ball in the following) 
on the dynamics of a coherently oscillating flat direction scalar 
$\Phi$, we will introduce a single Gaussian S-ball solution into the 
$\Phi$ scalar field equation. As discussed above, this is justified if 
the seperation of the S-balls is sufficiently 
large compared with their radius $R$ ($\delta x \gae 10 R$), in which case
there will be an extreme exponential suppression of the contribution of the 
other S-balls to the inflaton field 
in the vicinity of a given S-ball. In the space between the S-balls 
the extreme suppression will result in effectively no inflaton field or energy density.

    We will consider the case where the interaction ${\cal L}_{int}$
  comes purely from the 
derivative terms in \eq{e9}. This is exactly true for the case where
the $S$ scalar potential is assumed to come purely from soft SUSY 
breaking terms and 
radiative corrections, such that there is no superpotential for $S$. More generally, the
 effect of a 
superpotential will be to introduce a potential term
 into the inflaton F-term which will have at most the same magnitude as the 
derivative terms in the S-ball solution. Thus we expect similar
 results in the more general 
but model dependent case where there is a superpotential for the inflaton. 

      Since we are interested in the effect of the S-ball on a coherently 
oscillating flat direction condensate, we will consider a solution for 
$\Phi$ in the presence of an S-ball of the form, 
\be{e9a} \Phi(r,t) = \frac{\phi(r)}{\sqrt{2}}\sin(m_{\phi}t)    ~,\ee
 where $\phi(r)$ is a space-dependent 
amplitude for the coherent oscillation in the presence of the S-ball
 and $r$ is the radius 
from the S-ball centre. The $\Phi$ 
scalar field equation is then 
\be{e10} \ddot{\Phi} - \vec{\nabla}^{2}\Phi
 = - \left( m_{\phi}^{2}\Phi - \frac{\lambda}{M^{2}}
f(S) \Phi \right)    ~,\ee 
where 
\be{e11} f(S) = |\dot{S}|^{2} + |\vec{\nabla} S|^{2}         ~,\ee
and where as discussed above we consider the flat space scalar field equation. 
With $S$ given by the S-ball solution, \eq{e3}, $f(S)$ becomes 
\be{e12} f(S) = \left( \omega^{2} + \frac{4 r^{2}}{R^{4}} \right) 
\frac{s_{o}^{2}}{2}e^{-2 r^{2}/R^{2}}  
 ~.\ee
Substituting the coherently oscillating $\Phi$ solution, \eq{e9a}, into \eq{e10} 
then gives the equation for the flat direction scalar oscillation amplitude in the
 background of an S-ball,
\be{e13}    \phi^{''}(r) + \frac{2 \phi^{'}(r)}{r}  = -\frac{\lambda}{M^{2}} 
\left( \omega^{2} + \frac{4 r^{2}}{R^{4}} \right) 
\frac{s_{o}^{2}}{2}e^{-2 r^{2}/R^{2}}  \phi(r)
~,\ee
where $'$ denotes differentiation with respect to $r$. 

      We will look for a solution which is valid at $r/R \ll 1$, 
since the greatest effect of the inflaton energy density on the amplitude of coherent 
oscillations will be found at the centre of the S-ball. Suppose we consider 
$\phi(r) = \phi_{o}g_{s_{o}}(r)$, where $s_{o}$ is the value of the 
inflaton field at the centre of the S-ball. Far from the S-ball 
($r/R >> 1$) we expect that the 
function $g_{s_{o}}(r) \rightarrow 1$ if $\phi_{o}$ is the amplitude of the coherent 
oscillations in the absence of the S-ball. This is reasonable both physically and 
by inspection of the right hand side of \eq{e13}, which rapidly
 tends to zero as $r/R$ becomes 
large compared with 1, such that $\phi = \phi_{o}$ (constant)
 becomes a solution at large $r$. 
In addition, as $s_{o} \rightarrow 0$ 
the function $g_{s_{o}}(r) \rightarrow 1$ $\forall r$, since there is no
 S-ball in this limit. 
Therefore if we obtain a solution $\phi(r)$ valid at small $r$ then in order to relate it to 
the coherent oscillation amplitude far outside the S-ball we need
 only identify $g_{s_{o}}(r)$ 
by taking $s_{o} \rightarrow 0$ and setting $\phi(r) = \phi_{o}$ in this limit. 

  To find a solution valid at $r/R < 1$, 
we first change variable to $y = \log(\phi)$. Then \eq{e13} becomes 
\be{e14} y^{''} + y^{'\;2} + \frac{2 y^{'}}{r}   = -\frac{\lambda}{M^{2}} 
\left( \omega^{2} + \frac{4 r^{2}}{R^{4}} \right)
 \frac{s_{o}^{2}}{2}e^{-2 r^{2}/R^{2}}  ~.\ee
We then look for a solution of the form
\be{e15} y = A e^{-2r^{2}/R^{2}} + C    ~.\ee
Substituting this into \eq{e13}, the left hand side becomes
\be{e16} y^{''} + y^{'\;2} + \frac{2 y^{'}}{r}   
\equiv  \left(-\frac{12}{R^{2}}  
+  \frac{16 r^{2}}{R^{4}}\right) A e^{-2r^{2}/R^{2}} + 
\frac{16 r^{2}}{R^{4}} A^{2} e^{-4r^{2}/R^{2}}
~.\ee
Thus as $r/R \rightarrow 0$, 
\be{e17} y^{''} + y^{'\;2} + \frac{2 y^{'}}{r}  
\rightarrow  - \frac{12}{R^{2}}A e^{-2r^{2}/R^{2}} ~.\ee
This has the same form as the right hand side of 
\eq{e14} in the limit $r/R \ll \omega R /2$. Thus 
a solution valid for small $r/R$   
is 
\be{e18} \phi(r) = e^{y} =  e^{C} \exp\left( A e^{-2r^{2}/R^{2}} \right)   ~,\ee
where
\be{e19}  A = \frac{\lambda \omega^{2} R^{2} s_{o}^{2}}{24 M^{2}}   ~.\ee 
We see that as $s_{o} \rightarrow 0$, $\phi(r) \rightarrow e^{C}$. Therefore we 
set $\phi_{o} = e^{C}$, such that 
\be{e19a} \phi(r) = \phi_{o} \exp\left( 
\frac{\lambda \omega^{2} R^{2} s_{o}^{2} e^{-2r^{2}/R^{2}}}{24 M^{2}} 
  \right)  ~.\ee 
 Thus at the centre of the S-ball the 
amplitude of the coherently oscillating flat direction 
scalar as a function of $s_{o}$ is given by
\be{e20} \phi(r = 0) =  \phi_{o}\exp(A) \equiv 
\phi_{o}\exp\left(\frac{\lambda \omega^{2} R^{2} s_{o}^{2}}{24 M^{2}}  
\right)  ~.\ee

     From \eq{e20} we see that a significant change of the $\Phi$ amplitude at $r=0$ 
relative to its value in the absence of the S-ball, $\phi_{o}$,
 is possible only if $A \gae 1$. This requires that 
\be{e21} \frac{s_{o}}{M} \gae \sqrt{\frac{24}{\left|\lambda\right|}}
 \frac{1}{\omega R}    ~.\ee 
Since $\omega$ and $R$ for a Gaussian S-ball are typically determined by the 
same dynamical mass scale (the $S$ mass), $\omega R$ is expected 
to be not very much larger than 1. Therefore 
unless $s_{o}$ is close to the Planck scale, the S-ball will typically have little 
effect on the 
amplitude of flat direction coherent oscillations. Thus in the case of a NTS-dominated 
Universe the flat direction coherent oscillations are essentially unaltered from the case 
where there is no correction to the flat direction mass squared 
term i.e. the oscillations are 
{\it equivalent} to the case $c= 0$ of a homogeneous post-inflation Universe.  

              At first sight this result is surprising. It would seem that
 the large inflaton energy
 density and F-term would induce a large mass squared term
 in the vicinity of the S-ball,
 much larger than the order $H^{2}$ correction expected
 in the case of a homogeneous 
post-inflation Universe.
The reason that the S-ball has little effect on the coherent
 oscillations of the flat direction scalar is that as the flat direction
oscillation amplitude distorts under the influence of the induced mass squared term 
in the vicinity of the S-ball, the 
gradient energy of the now space-dependent amplitude
counteracts the effect of the 
mass squared term. It happens that the energy density
of the flat direction scalar is minimized 
when the distortion in the amplitude is negligibly small. 
To see this physically, consider the case where $\lambda > 0$ in \eq{e9},
 corresponding to a negative mass squared term for
 the flat direction scalar in the vicinity of the S-ball.
The contribution of the induced mass squared
 term $m_{eff}^{2} \approx -\rho_{S}/M^{2}$
in the vicinity of the S-ball (energy density $\approx \rho_{S}$)
to the energy density of the coherently oscillating flat direction scalar 
 is then, 
\be{add1} \rho_{m_{eff}} \approx  m_{eff}^{2} (\phi_{o}
 + \Delta \phi_{o})^{2} \approx - 
\frac{\rho_{S}}{M^{2}} (\phi_{o} + \Delta \phi_{o})^{2}  \approx 
- \frac{s_{o}^{2}}{M^{2} R^{2}} (\phi_{o} + \Delta \phi_{o})^{2}    ~,\ee
where the energy density inside the S-ball is expected to be of the
 order of $s_{o}^{2}/R^{2}$.
Here $\phi_{o}$ is the amplitude in the absence of the S-ball and $\Delta \phi_{o}$ is 
the change in the amplitude in the vicinity of the S-ball. 
The negative mass squared term 
will cause the amplitude of the coherent oscillations to increase in
 the vicinity of the S-ball, resulting in a gradient energy term $
\rho_{grad} \approx (\Delta \phi_{o}/R)^{2}$. 
Thus the energy density as a function of $\Delta \phi_{o}$ in the vicinity of the 
S-ball is given by 
\be{add2} \rho(\Delta \phi_{o}) \approx \rho_{m_{eff}} + \rho_{grad}
 = - \frac{s_{o}^{2} \left(\phi_{o} 
+ \Delta \phi_{o}\right)^{2}}{M^{2}R^{2}}
 + \frac{\Delta \phi_{o}^{2}}{R^{2}}   ~.\ee
This is minimized at 
\be{add3} \frac{\Delta \phi_{o}}{\phi_{o}} \approx
 \frac{s_{o}^{2}}{M^{2}}   ~.\ee 
Therefore if $s_{o}$ is small compared with the Planck scale then 
the shift of flat direction 
oscillation amplitude is negligible, in agreement with \eq{e21}. 

     For the case of the S-ball associated with running mass inflation the 
condition \eq{e21} becomes 
\be{e22} \frac{s_{o}}{M} \gae\sqrt{ 
\frac{12}{\left|\lambda\right|}} \frac{|K|^{1/2}}{ \left(1 + 
2 
|K|\right)^{1/2}}   ~.\ee
The value of $s_{o}$ when the S-balls form is $s_{o} \approx 10^{-2} |K| M$,
 with $|K| \approx 0.01-0.1$ \cite{krun}. 
Thus in this model the condition \eq{e21} is not satisfied and so there will be
 no significant effect on the dynamics of a coherently oscillating flat
 direction scalar i.e. the dynamics are equivalent to the case  $c = 0$ of a conventional
 homogeneous post-inflation era. Similarly, if we 
consider the case of D-term hybrid inflation and assume that the value
 of $s_{o}$ is characterised by 
the value of the inflaton field at the end of inflation, $s_{o} 
\approx 10^{16} \GeV $ \cite{icf,dti}, then the flat direction
 scalar dynamics will again 
effectively correspond to the case $c = 0$.
 Therefore we expect that the flat direction dynamics
 during an NTS-dominated post-inflation 
era will typically be equivalent to the case $c = 0$ in a conventional 
homogeneous post-inflation cosmology. 

       The end of the NTS-dominated era is expected to occur
 via the decay of the NTS. The value of 
$H$ at reheating is then given by $H(T_{R}) \approx
 k_{T_{R}}T_{R}^{2}/M_{Pl}$, where $T_{R}$ 
is the reheating temperature and $k_{T_{R}} \approx 20$. After reheating the energy 
density will be homogeneous and conventional 
$cH^{2}$ corrections with $|c| \approx 1$ will apply. 
Assuming that $T_{R} \lae 
10^{8-9} \GeV$ in order to evade the thermal gravitino
 problem \cite{grav} implies that reheating 
will occur at $H(T_{R}) \lae 1 \GeV$. Therefore when the $cH^{2}$ terms switch 
on they will already 
be small compared with the gravity-mediated soft SUSY breaking 
mass squared terms  
in the flat direction potential (of the order of $m_{W}^{2}$) and so will 
play no role in the dynamics of the flat direction scalars. 

     In the above we have assumed that the flat direction condensate
 does not modify the 
Gaussian NTS solution. The interaction term \eq{e9}
 also contributes terms to the 
$S$ field equation, 
\be{23} {\cal L}_{int} = ... + \frac{\lambda}{M^{2}} 
\left( \partial_{\mu}\Phi^{\dagger} \partial^{\mu}\Phi \right) |S|^{2}  
\rightarrow \frac{\lambda}{2 M^{2}} m_{\phi}^{2}
 \phi_{o}^{2} \cos^{2}\left(m_{\phi}t\right)|S|^{2} 
~.\ee
The condition for the flat direction condensate to have no effect 
on the S-ball solution is that this term is small compared
 with $m_{S}^{2}|S|^{2}$. This requires that, 
 \be{e24} \frac{\phi_{o}}{M}
 < \frac{m_{S}}{\sqrt{\lambda}m_{\phi}}     ~.\ee
Since $m_{S} \gae m_{\phi}$ is expected (where $m_{\phi}
 \approx 100 \GeV$ is the SUSY breaking mass term), this
 will typically be satisfied for all 
$\phi_{o} \lae M$. 

\section{Consequences for Affleck-Dine Baryogenesis and SUSY Curvatons}

       One common application of SUSY flat direction dynamics is to Affleck-Dine 
(AD) baryogenesis \cite{ad}. In the conventional homogeneous
 case we expect $|c| \approx 
1$ after the end 
of inflation. If $c$ is positive ($c \approx 1$) then the effect of the positive order
 $H^{2}$ term will be to 
drive damped oscillations of the flat direction scalar, such that 
the amplitude of the AD scalar oscillations is strongly suppressed by the time 
the gravity-mediated SUSY breaking mass squared term comes to dominate the 
dynamics and the baryon asymmetry in the condensate is fixed. As a result, 
AD baryogenesis is effectively ruled out for $c \approx 1$. Therefore the possibility of 
AD baryogenesis depends upon the (typically unknown) sign of $c$. For $c 
\approx -1$, the flat direction field will roll away from zero until the effect of 
non-renormalizable terms in the scalar potential stabilises the field at the minimum of 
its potential. In this case the initial amplitude of the AD scalar oscillations
is fixed by the dimension of the non-renormalizable superpotential terms which 
lift the flat direction potential and introduce the B and CP violation
 necessary for baryogenesis 
\cite{drt,ad,bbb}. As a result, the
 baryon asymmetry is fixed by the reheating temperature $T_{R}$ and the dimension 
$d$ of the non-renormalizable superpotential term. For $d = 4$ the reheating 
temperature must be of order $10^{8} \GeV$ in order
 to generate the observed asymmetry, 
whilst for $d = 6$ the reheating temperature must be around $1 \GeV$. (Even values 
of $d$ are 
required if R-parity is conserved \cite{drt,bbb2}, as suggested by the absence of
 dangerous renormalizable B- and L-violating contributions
 to the MSSM superpotential \cite{mssm}.)
Thus in the case of a homogeneous post-inflation
 cosmology with $|c| \approx 1$, the 
Affleck-Dine mechanism can only function if $c$ is negative and 
the reheating temperature must then either be high (close to the gravitino upper 
limit) or low ($1 \GeV$ or less) if R-parity is conserved. 

          In contrast, in the NTS-dominated case we have
 effectively $c \approx 0$. Therefore, 
the amplitude of the AD scalar remains fixed until the
 gravity-mediated SUSY breaking mass squared term comes to dominate the 
dynamics, as in the original
 Affleck-Dine baryogenesis scenario \cite{ad}. Thus in the
 NTS-dominated Universe the AD 
mechanism can, at least in principle, always generate the 
baryon asymmetry i.e. the AD scalar is never damped to zero. In
 addition, the asymmetry is not purely 
determined by $d$ and $T_{R}$ as in the $|c| \approx 1$ case,
but also by the amplitude of the AD scalar at the end of inflation.  
This should allow a wider range of MSSM flat directions 
and reheating temperatures to be compatible with the observed 
asymmetry.

     More recently, it has been suggested that a coherently oscillating 
scalar field (a 'curvaton') could be the source of cosmological
 density perturbations \cite{curv,curv2,curv3,clist,kcurv,postma,snu,phase}. 
A SUSY flat direction scalar could serve as a curvaton if
 its coherent oscillations can dominate the Universe before
 they decay, which requires a 
sufficiently large initial amplitude. Similar considerations
 then apply as in the case of the 
AD mechanism. In a conventional homogeneous post-inflation
 era with $|c| \approx 1$, a positive value of $c$
 will cause the curvaton to rapidly damp away after
 inflation. In this case the flat direction scalar energy
 density will be too small to dominate 
the Universe before it decays. Thus $c \approx -1$
 is necessary. However, this case also
 has a possible problem. If the curvaton rolls to the minimum of its potential, as
determined by the negative mass squared term and non-renormalizable superpotential
 terms, then oscillations of the curvaton about this minimum will damp the quantum
 fluctuations which lead to the energy density
 perturbations \cite{phase}. This will then require a large 
expansion rate during inflation to produce a sufficiently
 large quantum fluctuation, which may not be consistent with
 small enough adiabatic energy density perturbations from
 inflaton quantum fluctuations \cite{phase}. Both of these
 problems may be avoided in the case of a
 NTS-dominated post-inflation era , since effectively $c = 0$ in this case.

            These examples make it clear that an
inhomogeneous post-inflation era would have significant consequnces for SUSY
 cosmology. 

\section{Conclusions} 

          We have considered the effect of a non-topological
soliton-dominated post-inflation 
era on the dynamics of a coherently oscillating SUSY flat direction scalar.
 We have shown that the 
flat direction dynamics during a NTS-dominated era are typically not affected by the
 Planck-suppressed interactions which would generate a $cH^{2}$ mass squared
 term in the case of a conventional post-inflation cosmology. 
Thus during the
 NTS-dominated era we expect that the flat direction dynamics will 
effectively be equivalent to the case $c = 0$ of a homogeneous post-inflation era.

                 We have considered some 
consequences of a NTS-dominated era for 
SUSY cosmology. 
 Affleck-Dine baryogenesis will occur as in the orginal 
scenario with no order $H^{2}$ corrections. As a result, AD 
baryogenesis will become a more general possibility, since there will be no positive 
$H^{2}$ correction to drive the AD scalar
amplitude to zero before the asymmetry forms. In addition, the initial expectation value
 of the AD field is not fixed by the dimension of the terms lifting the flat direction,
 allowing a wider range of flat direction and reheating temperature 
to generate the observed baryon asymmetry. 
The dynamics of a curvaton in a NTS-dominated Universe
will be similarly modified, allowing a SUSY curvaton to have a large initial amplitude
 without quantum fluctuations of the curvaton being damped 
by oscillations about the minimum of the curvaton potential.

\end{document}